\documentclass[11pt]{article}

\usepackage{amssymb}
\usepackage[totalwidth=460truept,totalheight=600truept]{geometry}
\usepackage{latexsym,amssymb,amsmath,graphicx}
\usepackage[T1]{fontenc}

\def\theequation{\arabic{section}.\arabic{equation}}

\renewcommand{\theequation}{\thesection.\arabic{equation}}
\global\arraycolsep=1truept

\begin{document}

\hfill IFUP-TH 2009/20

\vskip 1.4truecm

\begin{center}
{\huge \textbf{Renormalization Of High-Energy }} \vskip .4truecm {\huge 
\textbf{Lorentz Violating QED}}

\vskip 1.5truecm

\textsl{{Damiano Anselmi and Martina Taiuti} }

\textit{Dipartimento di Fisica ``Enrico Fermi'', Universit\`{a} di Pisa, }

\textit{and INFN, Sezione di Pisa,}

\textit{Largo Pontecorvo 3, I-56127 Pisa, Italy, }

\textit{damiano.anselmi@df.unipi.it, martina.taiuti@df.unipi.it }

\textsl{\textit{\vskip 2truecm }}

\textsl{\textit{\textbf{{Abstract} }}}
\end{center}

{\small We study a QED extension that is unitary, CPT\ invariant and
super-renormalizable, but violates Lorentz symmetry at high energies, and
contains higher-dimension operators (LVQED). Divergent diagrams are only
one- and two-loop. We compute the one-loop renormalizations at high and low
energies and analyse the relation between them. It emerges that the
power-like divergences of the low-energy theory are multiplied by arbitrary
constants, inherited by the high-energy theory, and therefore can be set to
zero at no cost, bypassing the hierarchy problem.}

\vskip 1truecm

\vfill\eject

\section{Introduction}

\setcounter{equation}{0}

Experimental measurements and observations tell us that Lorentz symmetry is
one of the most precise symmetries in nature \cite{kostelecky}.
Nevertheless, the possibility that Lorentz symmetry might be violated at
high energies or very large distances has been widely investigated. From the
theoretical point of view, it is interesting to know that if Lorentz
symmetry is violated at high energies, vertices that are non-renormalizable
by power counting can become renormalizable by a modified power counting
criterion, which weights space and time differently \cite{halat}. In the
common perturbative framework, the theory remains unitary, local, polynomial
and causal.

Recently, a Lorentz violating CPT\ invariant Standard Model extension
inspired by this idea has been formulated \cite{lvsm,noh}. Its main property
is that it contains two scalar-two fermion vertices, as well as four fermion
vertices, at the fundamental level. In particular, four fermion vertices can
trigger a Nambu--Jona-Lasinio mechanism, that gives masses both to the
fermions and the gauge fields, even if the elementary Higgs boson is
suppressed \cite{noh}. In its simplest version, the scalarless model
schematically reads 
\begin{equation}
\mathcal{L}_{\mathrm{noH}}=\mathcal{L}_{Q}+\mathcal{L}_{\mathrm{{kin}f}%
}-\sum_{I=1}^{5}\frac{1}{\Lambda _{L}^{2}}g\bar{D}\bar{F}\,(\bar{\chi}_{I}%
\bar{\gamma}\chi _{I})+\frac{Y_{f}}{\Lambda _{L}^{2}}\bar{\chi}\chi \bar{\chi%
}\chi -\frac{g}{\Lambda _{L}^{2}}\bar{F}^{3},  \label{noH}
\end{equation}
where 
\begin{eqnarray*}
\mathcal{L}_{Q} &=&-\frac{1}{4}\sum_{G}\left( 2F_{\hat{\mu}\bar{\nu}%
}^{G}F_{{}}^{G\hspace{0.01in}\hat{\mu}\bar{\nu}}+F_{\bar{\mu}\bar{\nu}%
}^{G}\tau ^{G}(\bar{\Upsilon})F^{G\hspace{0.01in}\bar{\mu}\bar{\nu}}\right) ,
\\
\mathcal{L}_{\mathrm{{kin}f}} &=&\sum_{a,b=1}^{3}\sum_{I=1}^{5}\bar{\chi}%
_{I}^{a}\hspace{0.02in}i\left( \delta ^{ab}\hat{D}\!\!\!\!\slash -\frac{%
b_{0}^{Iab}}{\Lambda _{L}^{2}}{\bar{D}\!\!\!\!\slash}\,^{3}+b_{1}^{Iab}\bar{D%
}\!\!\!\!\slash \right) \chi _{I}^{b}.
\end{eqnarray*}
Hats are used to denote time components, bars to denote space components.
The field strengths are decomposed in $F_{\bar{\mu}\bar{\nu}}$, also denoted
with $\bar{F}$, and $F_{\hat{\mu}\bar{\nu}}$. Moreover, $\chi
_{1}^{a}=L^{a}=(\nu _{L}^{a},\ell _{L}^{a})$, $\chi
_{2}^{a}=Q_{L}^{a}=(u_{L}^{a},d_{L}^{a})$, $\chi _{3}^{a}=\ell _{R}^{a}$, $%
\chi _{4}^{a}=u_{R}^{a}$ and $\chi _{5}^{a}=d_{R}^{a}$, $\nu ^{a}=(\nu
_{e},\nu _{\mu },\nu _{\tau })$, $\ell ^{a}=(e,\mu ,\tau )$, $u^{a}=(u,c,t)$
and $d^{a}=(d,s,b)$. The sum $\sum_{G}$ is over the gauge groups $SU(3)_{c}$%
, $SU(2)_{L}$ and $U(1)_{Y}$, and the last three terms of (\ref{noH}) are
symbolic. Finally, $\bar{\Upsilon}\equiv -\bar{D}^{2}/\Lambda _{L}^{2}$,
where $\Lambda _{L}$ is the scale of Lorentz violation, and $\tau ^{G}$ are
polynomials of degree 2.

The weight of time is $-1$, the one of the space coordinates is $-$1/3, so
the weights of energy and momentum are 1 and 1/3, respectively. The theory
has weighted dimension 2, so the lagrangian contains only terms of weights $%
\leqslant 2$. The weight of the gauge couplings $g$ is 1/3. Gauge anomalies
cancel out exactly as in the Standard Model \cite{lvsm}. The ``boundary
conditions'' that ensure that Lorentz invariance is recovered at low
energies are that $b_{1}^{Iab}$ tend to $\delta ^{ab}$ and $\tau ^{G}$ tend
to 1. One such condition can be trivially fulfilled normalizing the space
coordinates $\bar{x}$.

The purpose of this paper is to begin a systematic investigation of the
renormalization of the model (\ref{noH}), starting from its electromagnetic
sector, which we dub LVQED. From the high-energy point of view, the most
important novelty is that the electric charge is super-renormalizable. Thus,
the simplest version of LVQED is asymptotically free, with a finite number
of divergent diagrams (at one and two loops).

The low-energy theory, which we dub lvQED, is obtained taking the limit $%
\Lambda _{L}\rightarrow \infty $, where the weighted power counting is
replaced by ordinary power counting. lvQED is a power-counting
renormalizable, but Lorentz violating, electrodynamics. Studying the
interpolation between the renormalizations of LVQED\ and lvQED, we show that
the power-like divergences of lvQED (expressed as powers of $\Lambda _{L}$)
are multiplied by arbitrary coefficients, inherited by the high-energy
theory. This is a very general property of high-energy Lorentz violating
theories, and holds also in the Lorentz violating Standard Model (\ref{noH})
and the other versions formulated in ref.s \cite{lvsm,noh}. If the
elementary Higgs field is present, the arbitrariness just mentioned can be
used to remove the hierarchy problem.

The paper is organized as follows. In section 2 we present the simplest
version of LVQED and quantize it using the functional integral. In section 3
we work out its one-loop renormalization. In section 4 we study its
low-energy limit and compare the renormalizations of LVQED\ and lvQED,
pointing out the arbitrariness multiplying the low-energy power-like
divergences. In section 5 we work out the one-loop renormalization of lvQED.
In section 6 we reconsider the hierarchy problem in the light of our
results. Section 7 contains our conclusions. In the appendices we collect
some details about the calculations.

\section{The theory}

\setcounter{equation}{0}

The simplest form of LVQED is 
\begin{eqnarray}
\mathcal{L} &=&-\frac{1}{2}F_{\hat{\mu}\bar{\nu}}F^{\hat{\mu}\bar{\nu}}-%
\frac{1}{4}F_{\bar{\mu}\bar{\nu}}\left( \tau _{2}-\tau _{1}\frac{\bar{%
\partial}^{2}}{\Lambda _{L}^{2}}+\tau _{0}\frac{(-\bar{\partial}^{2})^{2}}{%
\Lambda _{L}^{4}}\right) F^{\bar{\mu}\bar{\nu}}  \nonumber \\
&&+\bar{\psi}\left( i\hat{D}\!\!\!\!\slash +\frac{ib_{0}}{\Lambda _{L}^{2}}{%
\bar{D}\!\!\!\!\slash}\,^{3}+ib_{1}\bar{D}\!\!\!\!\slash -m-\frac{b^{\prime }%
}{\Lambda _{L}}{\bar{D}\!\!\!\!\slash}\,^{2}\right) \psi  \label{mink} \\
&&+\frac{e}{\Lambda _{L}}\bar{\psi}\hspace{0.02in}\left( b^{\prime \prime
}\sigma _{\bar{\mu}\bar{\nu}}F^{\bar{\mu}\bar{\nu}}+\frac{b_{0}^{\prime }}{%
\Lambda _{L}}\gamma _{\bar{\mu}}\partial _{\bar{\nu}}F^{\bar{\mu}\bar{\nu}%
}\right) \psi +ie\frac{b_{0}^{\prime \prime }}{\Lambda _{L}^{2}}F^{\bar{\mu}%
\bar{\nu}}\left( \bar{\psi}\gamma _{\bar{\mu}}\hspace{0.02in}\frac{%
\overleftrightarrow{{\bar{D}}}_{\bar{\nu}}}{2}\psi \right) ,  \nonumber
\end{eqnarray}
where the covariant derivative reads $D_{\mu }=\partial _{\mu }+ieA_{\mu }$
and $\sigma _{\mu \nu }=-i[\gamma _{\mu },\gamma _{\nu }]/2$. The lagrangian
(\ref{mink}) is obtained including the smallest set of terms that are closed
under renormalization, together with their ``non-minimal'' and more relevant
partners. For example, since $\bar{\psi}\hspace{0.02in}{\bar{D}\!\!\!\!\slash%
}\,^{3}\psi $ must be present (to ensure that the fermion propagator falls
off sufficiently rapidly in the space directions), so are the terms $\sim 
\bar{\psi}{\bar{D}\!\!\!\!\slash}\,^{i}\bar{F}\psi $, $i\leqslant 1$, and $%
\bar{\psi}{\bar{D}\!\!\!\!\slash}\,^{i}\psi $, with $i<3$.

To study renormalization, it is convenient to turn to Euclidean space. In
our models the Wick rotation is straightforward because the time-derivative
structure is the same as in ordinary quantum field theories, and therefore
also the pole structure of propagators and amplitudes. In ref. \cite
{confnolor} it was shown that the K\"{a}llen-Lehman spectral decomposition,
the cutting equations, as well as the unitarity relation and Bogoliubov's
causality \cite{bogo}\footnote{%
The most general formulation of Bogoliubov's causality is an identity
satisfied by the $S$ matrix, which does not require light cones, but just
past and future. An elegant proof that can be easily generalized to Lorentz
violating theories is given in \cite{diagra}.}, can be generalized to our
types of Lorentz violating theories. The theorem of locality of counterterms
ensures that the renormalization constants are the same before and after the
Wick rotation.

The Euclidean lagrangian reads 
\begin{eqnarray}
\mathcal{L}_{\mathrm{E}} &=&\frac{1}{2}F_{\hat{\mu}\bar{\nu}}F_{\hat{\mu}%
\bar{\nu}}+\frac{1}{4}F_{\bar{\mu}\bar{\nu}}\left( \tau _{2}-\tau _{1}\frac{%
\bar{\partial}^{2}}{\Lambda _{L}^{2}}+\tau _{0}\frac{(-\bar{\partial}%
^{2})^{2}}{\Lambda _{L}^{4}}\right) F_{\bar{\mu}\bar{\nu}}  \nonumber \\
&&+\bar{\psi}\left( \hat{D}\!\!\!\!\slash -\frac{b_{0}}{\Lambda _{L}^{2}}{%
\bar{D}\!\!\!\!\slash}\,^{3}+b_{1}\bar{D}\!\!\!\!\slash +m-\frac{b^{\prime }%
}{\Lambda _{L}}{\bar{D}\!\!\!\!\slash}\,^{2}\right) \psi  \label{le} \\
&&+\frac{e}{\Lambda _{L}}\bar{\psi}\hspace{0.02in}\left( b^{\prime \prime
}\sigma _{\bar{\mu}\bar{\nu}}F_{\bar{\mu}\bar{\nu}}+\frac{ib_{0}^{\prime }}{%
\Lambda _{L}}\gamma _{\bar{\mu}}\partial _{\bar{\nu}}F_{\bar{\mu}\bar{\nu}%
}\right) \psi +e\frac{b_{0}^{\prime \prime }}{\Lambda _{L}^{2}}F_{\bar{\mu}%
\bar{\nu}}\left( \bar{\psi}\hspace{0.02in}\gamma _{\bar{\mu}}\frac{%
\overleftrightarrow{{\bar{D}}}_{\bar{\nu}}}{2}\psi \right) ,  \nonumber
\end{eqnarray}
and the covariant derivative keeps its form $D_{\mu }=\partial _{\mu
}+ieA_{\mu }$.

To ensure a positive definite bosonic sector we must assume 
\[
\tau _{0}>0,\qquad \tau _{2}>0,\qquad \tau _{1}^{2}\leqslant 4\tau _{0}\tau
_{2}. 
\]

\paragraph{Gauge-fixing and propagators}

The BRST\ symmetry coincides with the one of Lorentz invariant QED, namely 
\[
sA_{\mu }=\partial _{\mu }C,\qquad sC=0,\qquad s\bar{C}=B,\qquad sB=0, 
\]
\noindent where $B$ is a Lagrange multiplier. We choose the ``Feynman''
gauge-fixing lagrangian 
\begin{equation}
\mathcal{L}_{\mathrm{GF}}=s\left[ \bar{C}\left( -\frac{1}{2}\tau (-\bar{%
\partial}^{2}/\Lambda _{L}^{2})B+\hat{\partial}\hat{A}+\tau (-\bar{\partial}%
^{2}/\Lambda _{L}^{2})\bar{\partial}\cdot \bar{A}\right) \right] ,
\label{lgf}
\end{equation}
where $\tau $ is the polynomial $\tau (x)=\tau _{2}+\tau _{1}x+\tau
_{0}x^{2} $. Integrating $B$ out we find 
\begin{equation}
\mathcal{L}_{\mathrm{GF}}\rightarrow (\hat{\partial}\hat{A}+\tau \bar{%
\partial}\cdot \bar{A})\frac{1}{2\tau }(\hat{\partial}\hat{A}+\tau \bar{%
\partial}\cdot \bar{A})-\bar{C}(\hat{\partial}^{2}+\tau \bar{\partial}%
^{2})C\rightarrow (\hat{\partial}\hat{A}+\tau \bar{\partial}\cdot \bar{A})%
\frac{1}{2\tau }(\hat{\partial}\hat{A}+\tau \bar{\partial}\cdot \bar{A}).
\label{nonlo}
\end{equation}
As in usual QED, the ghosts decouple, so from now on we ignore them. Observe
that (\ref{nonlo}) is strictly speaking non-local, since $\tau $ appears in
the denominator. However, this is not a problem, since (\ref{lgf}) is local
and the propagators are well-behaved. The photon propagator reads 
\begin{eqnarray*}
\langle \hat{A}(k)\hspace{0.01in}\hat{A}(-k)\rangle &=&\frac{\tau (\bar{k}%
^{2}/\Lambda _{L}^{2})}{\hat{k}^{2}+\bar{k}^{2}\tau (\bar{k}^{2}/\Lambda
_{L}^{2})},\qquad \langle \hat{A}(k)\hspace{0.01in}\bar{A}_{\bar{\mu}%
}(-k)\rangle =0, \\
\langle \bar{A}_{\bar{\mu}}(k)\hspace{0.01in}\bar{A}_{\bar{\nu}}(-k)\rangle
&=&\frac{\delta _{\bar{\mu}\bar{\nu}}}{\hat{k}^{2}+\bar{k}^{2}\tau (\bar{k}%
^{2}/\Lambda _{L}^{2})},
\end{eqnarray*}
while the electron propagator is a bit more involved: 
\[
\langle \psi (p)\hspace{0.01in}\bar{\psi}(-p)\rangle =\frac{-i\hat{p}\!\!\!%
\slash -i\bar{p}\!\!\!\slash M+N}{\hat{p}^{2}+\bar{p}^{2}M^{2}+N^{2}}, 
\]
where 
\[
M=b_{1}+\frac{b_{0}}{\Lambda _{L}^{2}}\bar{p}^{2},\qquad N=m+\frac{b^{\prime
}}{\Lambda _{L}}\bar{p}^{2}. 
\]

\paragraph{Propagating degrees of freedom}

The propagating degrees of freedom can be exhibited in the ``Coulomb''
gauge, choosing 
\[
\mathcal{L}_{\mathrm{GF}}=s\left[ \bar{C}\left( -\frac{\lambda }{2}B+\bar{%
\partial}\cdot \bar{A}\right) \right] \rightarrow \frac{1}{2\lambda }(\bar{%
\partial}\cdot \bar{A})^{2}-\bar{C}\bar{\partial}^{2}C. 
\]
The ghosts are non-propagating, since their two-point function does not
contain poles. Instead, the photon propagator in the Coulomb gauge reads 
\begin{eqnarray*}
\langle \hat{A}(k)\hspace{0.01in}\hat{A}(-k)\rangle &=&\frac{1}{\bar{k}^{2}}+%
\frac{\lambda \hat{k}^{2}}{(\bar{k}^{2})^{2}},\qquad \langle \hat{A}(k)%
\hspace{0.01in}\bar{A}(-k)\rangle =\frac{\lambda \hat{k}\bar{k}}{(\bar{k}%
^{2})^{2}}, \\
\langle \bar{A}(k)\hspace{0.01in}\bar{A}(-k)\rangle &=&\frac{1}{\hat{k}%
^{2}+\tau \bar{k}^{2}}\left( \bar{\delta}-\frac{\bar{k}\bar{k}}{\bar{k}^{2}}%
\right) +\frac{\lambda \bar{k}\bar{k}}{(\bar{k}^{2})^{2}}.
\end{eqnarray*}
Writing $k^{\mu }=(iE,\mathbf{k})$ and studying the poles, we see that the
propagating degrees of freedom are two, as expected, with the dispersion
relation 
\[
E=|\bar{k}|\sqrt{\tau _{2}+\tau _{1}\frac{\bar{k}^{2}}{\Lambda _{L}^{2}}%
+\tau _{0}\frac{(\bar{k}^{2})^{2}}{\Lambda _{L}^{4}}}. 
\]

As usual, the Coulomb gauge exhibits unitarity, the Feynman gauge exhibits
renormalizability. Gauge independence ensures that the physical correlation
functions are both unitary and renormalizable.

\paragraph{Regularization}

A convenient all-order regularization technique is \cite{lvsm} a combination
of a higher-derivative regularization \textit{\`{a} la} Slavnov \cite
{slavnov}, for diagrams with two and more loops, combined with the
dimensional regularization for one-loop diagrams. Thus, for our present
interests, which are restricted to one-loop integrals, we just need the
dimensional regularization. In principle, we should dimensionally continue
both time and space. However, the calculations of this paper are all
convergent in the hatted direction, so we just need to continue space to $%
3-\varepsilon _{2}$ dimensions, with $\varepsilon _{2}$ complex.

As usual, to renormalize the high-energy theory, it is necessary to
introduce a dynamical scale $\mu $, which we define to have weight one and
dimension one.

\paragraph{Weights and dimensions}

We list here the weights of fields and parameters, denoted with square
brackets. In the physical limit ($\varepsilon _{2}=0$) we have 
\begin{eqnarray}
\lbrack \mu ] &=&[m]=[\hat{\partial}]=1,\qquad [\bar{\partial}]=\frac{1}{3}%
,\qquad [\hat{A}]=\frac{2}{3},\quad \quad [\bar{A}]=0,\quad \quad [\psi ]=%
\frac{1}{2},\qquad [\tau _{2}]=\frac{4}{3},  \nonumber \\
\lbrack b_{0}] &=&[b_{0}^{\prime }]=[b_{0}^{\prime \prime }]=[\tau
_{0}]=[\Lambda _{L}]=0,\qquad [e]=[b^{\prime }]=[b^{\prime \prime }]=\frac{1%
}{3},\qquad [b_{1}]=[\tau _{1}]=\frac{2}{3}.  \label{wei}
\end{eqnarray}
Thus, the electric and magnetic fields have weights 1 and 1/3, respectively (%
$[F_{\hat{\mu}\bar{\nu}}]=1$, $[F_{\bar{\mu}\bar{\nu}}]=1/3$). After
dimensional continuation, all quantities keep their weights unchanged,
except for the fields and the electric charge, which acquire the weights 
\begin{equation}
\lbrack \hat{A}]=\frac{2}{3}-\frac{\varepsilon _{2}}{6},\quad \quad [\bar{A}%
]=-\frac{\varepsilon _{2}}{6},\quad \quad [\psi ]=\frac{1}{2}-\frac{%
\varepsilon _{2}}{6},\qquad [e]=\frac{1}{3}+\frac{\varepsilon _{2}}{6}.
\label{bw}
\end{equation}

The dimensions of fields in units of mass are just the usual ones. All
parameters are dimensionless, except for $\Lambda _{L}$ and $\mu $, which
have dimension one.

For the purposes of renormalization, the weightful parameters $e$, $%
b^{\prime }$, $b^{\prime \prime }$, $b_{1}$, $\tau _{1}$, $\tau _{2}$ and $m$
can be treated perturbatively, since the divergent parts of diagrams depend
polynomially on them. They can be understood as parameters multiplying
``two-leg vertices''. Intermediate infrared problems can be avoided
introducing a fictitious mass $\delta $ in the denominators, which must be
set to zero after the calculation of the divergent part (which is also
polynomial in $\delta $). Of course this trick cannot be used if we want to
calculate the finite parts of correlation functions. Thus, we use the
propagators 
\[
\langle \hat{A}(k)\hspace{0.01in}\hat{A}(-k)\rangle =\tau _{0}\frac{(\bar{k}%
^{2})^{2}}{\Lambda _{L}^{4}}\frac{1}{\hat{k}^{2}+\tau _{0}\frac{(\bar{k}%
^{2})^{3}}{\Lambda _{L}^{4}}+\delta ^{2}},\qquad \langle \bar{A}_{\bar{\mu}%
}(k)\hspace{0.01in}\bar{A}_{\bar{\nu}}(-k)\rangle =\frac{\delta _{\bar{\mu}%
\bar{\nu}}}{\hat{k}^{2}+\tau _{0}\frac{(\bar{k}^{2})^{3}}{\Lambda _{L}^{4}}%
+\delta ^{2}}, 
\]
and $\langle \hat{A}(k)\hspace{0.01in}\bar{A}_{\bar{\mu}}(-k)\rangle =0$ for
the photon and 
\[
\langle \psi (p)\hspace{0.01in}\bar{\psi}(-p)\rangle =\frac{-i\hat{p}\!\!\!%
\slash -ib_{0}\frac{\bar{p}^{2}}{\Lambda _{L}^{2}}\bar{p}\!\!\!\slash }{\hat{%
p}^{2}+b_{0}^{2}\frac{(\bar{p}^{2})^{3}}{\Lambda _{L}^{4}}+\delta ^{2}} 
\]
for the electron. Using this trick, we can expand diagrams both in the
external momenta and in the weightful parameters. At the end all one-loop
divergences can be reduced to the divergent part of one integral, reported
in appendix A.

\paragraph{Bare and regularized theories}

If the fields and parameters of (\ref{le}) are interpreted as bare, (\ref{le}%
) becomes the bare lagrangian. The weights of bare fields, renormalized
fields and bare parameters are those of (\ref{bw}), while the weights of
renormalized parameters are given in (\ref{wei}).

We know that there are no wave-function renormalization constants (because
the theory is super-renormalizable), so bare and renormalized fields
coincide. By the Ward identity, which is easy to prove, the electric charge
is not renormalized either. Moreover, we have parametrized (\ref{le}) so
that each vertex carries a power of $e$ equal to the number of its legs
minus 2. Then, it is simple to prove that each loop carries an additional
factor $e^{2}$, which has weight 2/3. This ensures that no parameter with
weight $\leqslant 1/3$ can have a non-trivial renormalization.

The only nontrivial relations among bare and renormalized parameter can be
expressed as 
\begin{eqnarray}
e_{\mathrm{B}} &=&e_{\mathrm{R}}\mu ^{\varepsilon _{2}/6}\Lambda
_{L}{}^{\varepsilon _{2}/3},\qquad m_{\mathrm{B}}=m_{\mathrm{R}}+\delta
^{(1)}m_{\mathrm{R}},\qquad b_{1\mathrm{B}}=b_{1\mathrm{R}}+\delta ^{(1)}b_{1%
\mathrm{R}},  \nonumber \\
\tau _{2\mathrm{B}} &=&\tau _{2\mathrm{R}}+\delta ^{(1)}\tau _{2\mathrm{R}%
}+\delta ^{(2)}\tau _{2\mathrm{R}},\qquad \tau _{1\mathrm{B}}=\tau _{1%
\mathrm{R}}+\delta ^{(1)}\tau _{1\mathrm{R}},  \label{cup}
\end{eqnarray}
where $\delta ^{(1)}$ and $\delta ^{(2)}$ denote the one- and two-loop
contributions, respectively.

The relations (\ref{cup}), the first one in particular, are obtained
matching the dimensions and weights of bare and renormalized parameters,
recalling that $\Lambda _{L}$ is weightless, while the dynamical scale $\mu $
has weight 1. Because two-loop diagrams carry a factor $e^{4}$, only $\tau
_{2}$ can have a non-trivial two-loop renormalization. Finally, it is
important to bear in mind that $\Lambda _{L}$ is not renormalized, since it
is a redundant parameter.

\section{High-energy renormalization}

\setcounter{equation}{0}

In this section we study the one-loop renormalization of LVQED. The one-loop
divergent diagrams are depicted in figure 1, where the double curly line
denotes $\hat{A}$ and the simple curly line denotes $\bar{A}$.

\begin{figure}[tbp]
\centerline{\includegraphics[width=5in,height=2.5in]{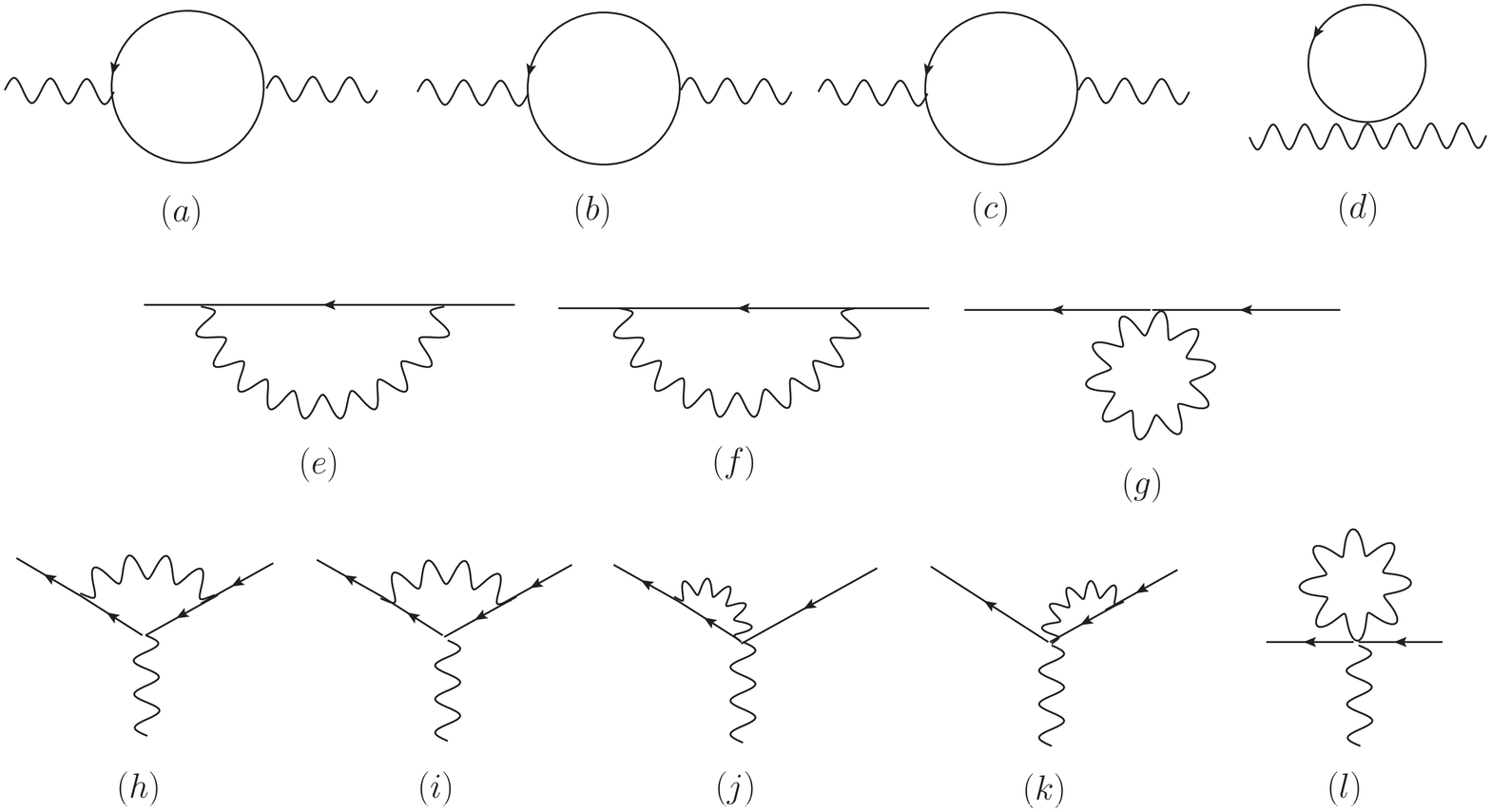}}
\caption{One-loop divergent diagrams}
\end{figure}
By weighted power counting, if diagram (a) were divergent it would produce a
mass term $e^{2}\hat{A}^{2}$. However, the divergent part of diagram (a) is
proportional to 
\[
4e^{2}\int_{-\infty }^{+\infty }\frac{\mathrm{d}\hat{p}}{2\pi }\int \frac{%
\mathrm{d}^{3-\varepsilon _{2}}\bar{p}}{(2\pi )^{3}}\frac{\hat{p}%
^{2}-b_{0}^{2}\frac{(\bar{p}^{2})^{3}}{\Lambda _{L}^{4}}}{\left( \hat{p}%
^{2}+b_{0}^{2}\frac{(\bar{p}^{2})^{3}}{\Lambda _{L}^{4}}+\delta ^{2}\right)
^{2}}, 
\]
so it vanishes because of the identity \cite{noh} 
\[
\int_{-\infty }^{+\infty }\frac{\mathrm{d}\hat{p}}{2\pi }\frac{\hat{p}^{2}-x%
}{(\hat{p}^{\hspace{0.01in}2}+x)^{2}}=0,\qquad x>0. 
\]
Diagram (b) vanishes because its integrand is odd in $\hat{p}$. All other
diagrams are non-trivial.

The calculation of one-loop divergences gives the counterterms 
\[
\Delta _{1}\mathcal{L}_{\mathrm{E}}=\frac{\Delta _{1}\tau _{2}}{4\varepsilon
_{2}}F_{\bar{\mu}\bar{\nu}}^{2}-\frac{\Delta _{1}\tau _{1}}{4\varepsilon _{2}%
}F_{\bar{\mu}\bar{\nu}}\frac{\bar{\partial}^{2}}{\Lambda _{L}^{2}}F_{\bar{\mu%
}\bar{\nu}}+\frac{1}{\varepsilon _{2}}\bar{\psi}\left( \Delta _{1}b_{1}\bar{D%
}\!\!\!\!\slash +\Delta _{1}m\right) \psi , 
\]
where 
\begin{eqnarray}
\Delta _{1}\tau _{2} &=&\frac{s_{0}e^{2}}{6\pi ^{2}}\left( -b_{1}-4\frac{%
b_{0}^{\prime \prime \hspace{0.01in}2}b_{1}}{b_{0}^{2}}-\frac{1}{2}\frac{%
b^{\prime \hspace{0.01in}2}}{b_{0}}-2\frac{b_{0}^{\prime \prime \hspace{%
0.01in}2}b^{\prime \hspace{0.01in}2}}{b_{0}^{3}}-12\frac{b^{\prime
}b^{\prime \prime }}{b_{0}}+8\frac{b^{\prime \prime \hspace{0.01in}2}}{b_{0}}%
\right) ,  \nonumber \\
\Delta _{1}\tau _{1} &=&-\frac{e^{2}|b_{0}|}{6\pi ^{2}}\left( \frac{3}{10}+2%
\frac{b_{0}^{\prime }}{b_{0}}-4\frac{b_{0}^{\prime \ 2}}{b_{0}^{2}}+\frac{11%
}{5}\frac{b_{0}^{\prime \prime \hspace{0.01in}2}}{b_{0}^{2}}\right) , 
\nonumber \\
\Delta _{1}b_{1} &=&\frac{e^{2}}{3\pi ^{2}(|b_{0}|+\sqrt{\tau _{0}})^{2}}%
\left( -\frac{9}{2}s_{0}b_{0}^{2}+|b_{0}|b_{0}^{\prime }-s_{0}b_{0}^{\prime
\prime \hspace{0.01in}2}+4\frac{b_{0}^{2}b_{0}^{\prime }}{\sqrt{\tau _{0}}}%
\right.  \label{formu} \\
&&\left. \qquad \qquad \qquad \qquad +\frac{3}{2}\frac{b_{0}b_{0}^{\prime 
\hspace{0.01in}2}}{\sqrt{\tau _{0}}}-\frac{5}{8}\frac{b_{0}b_{0}^{\prime
\prime \hspace{0.01in}2}}{\sqrt{\tau _{0}}}-\frac{7}{2}b_{0}\sqrt{\tau _{0}}-%
\frac{1}{2}s_{0}\tau _{0}\right) ,  \nonumber \\
\Delta _{1}m &=&\frac{e^{2}\Lambda _{L}{}}{\pi ^{2}(|b_{0}|+\sqrt{\tau _{0}})%
}\left( -\frac{3}{4}b^{\prime }-\frac{b_{0}^{\prime \hspace{0.01in}%
2}b^{\prime }}{2|b_{0}|\sqrt{\tau _{0}}}-\frac{b_{0}^{\prime \prime \hspace{%
0.01in}2}b^{\prime }}{8|b_{0}|\sqrt{\tau _{0}}}+2\frac{|b_{0}|b^{\prime
\prime }}{\sqrt{\tau _{0}}}+2\frac{s_{0}b_{0}^{\prime }b^{\prime \prime }}{%
\sqrt{\tau _{0}}}-\frac{b^{\prime }\sqrt{\tau _{0}}}{4|b_{0}|}\right) , 
\nonumber
\end{eqnarray}
and $s_{0}=b_{0}/|b_{0}|$. The fact that the sets of counterterms $\bar{\psi}%
\bar{\partial}\!\!\!\slash \psi $ and $\bar{\psi}\bar{A}\!\!\!\slash \psi $
combine to reconstruct the gauge-invariant expression $\bar{\psi}\bar{D}%
\!\!\!\!\slash \psi $ is a check of our results.

With respect to formulas (\ref{cup}) we have $\delta ^{(1)}g_{\mathrm{R}%
}=(\Delta _{1}g)/\varepsilon _{2}$, where $g=\tau _{2}$, $\tau _{1}$, $b_{1}$
or $m$. Moreover, the first of (\ref{cup}) gives 
\[
\frac{e^{2}}{\varepsilon _{2}}=\frac{1}{\varepsilon _{2}}e_{\mathrm{B}%
}^{2}\mu ^{-\varepsilon _{2}/3}\Lambda _{L}{}^{-2\varepsilon _{2}/3},\qquad
\mu \frac{\mathrm{d}}{\mathrm{d}\mu }\left( \frac{e^{2}}{\varepsilon _{2}}%
\right) =-\frac{e^{2}}{3}, 
\]
\noindent so the one-loop beta functions are 
\[
\beta _{g}=\frac{1}{3}\Delta _{1}g. 
\]

\section{Relation between low-energy and high-energy divergences}

\setcounter{equation}{0}

In this section we study the renormalization of the low-energy theory and
its relation with the renormalization of the high-energy theory.

The low-energy limit of LVQED can be studied taking the limit $\Lambda
_{L}\rightarrow \infty $ in the physical correlation functions. It is
described by the lagrangian 
\begin{equation}
\mathcal{L}_{\mathrm{E}\text{\textrm{-}}\mathrm{low}}=\frac{1}{2}F_{\hat{\mu}%
\bar{\nu}}F_{\hat{\mu}\bar{\nu}}+\frac{\tau _{2}}{4}F_{\bar{\mu}\bar{\nu}}F_{%
\bar{\mu}\bar{\nu}}+\bar{\psi}\left( \hat{D}\!\!\!\!\slash +b_{1}\bar{D}%
\!\!\!\!\slash +m\right) \psi ,  \label{lelow}
\end{equation}
(in Euclidean space). We refer to this theory as lvQED. The low-energy
values of $\tau _{2}$ and $b_{1}$ have to be sufficiently close to 1 to have
agreement with experiments (see \cite{kostelecky}). Here, however, we are
interested in more theoretical aspects. Our goal is to compare the
renormalizations of LVQED and lvQED, and explain in detail how the
high-energy divergences in $1/\varepsilon _{2}$ combine with the $\Lambda
_{L}$-divergences to reproduce the low-energy results. We discover that the
low-energy power-like divergences are multiplied by arbitrary constants,
inherited by the high-energy theory. This makes the hierarchy problem
disappear.

Let us call the theory (\ref{le}), equipped with its
dimensional-regularization technique, LVQED$_{\varepsilon }$. From the
low-energy point of view, LVQED$_{\varepsilon }$ can be viewed as a
particular regularization of (\ref{lelow}) with a combination of two
cut-offs: the dimensional one and $\Lambda _{L}$.

Specifically, if $\Lambda _{L}$ is viewed as a cut-off, (\ref{le}) can be
understood as a (partial) regularization of (\ref{lelow}). The
regularization is then completed dimensionally continuing the space
dimensions to $3-\varepsilon _{2}$, with the prescription that the limit $%
\varepsilon _{2}\rightarrow 0$ be taken \textit{before} the limit $\Lambda
_{L}\rightarrow \infty $.

Recall that when two or more cut-offs are used to regularize a theory they
can be removed in any preferred order, up to a change of scheme. In a single
one-loop integral, the result can change at most by local terms, which are
possibly divergent. In higher-loop integrals the same conclusion holds when
the subdivergences are removed by appropriate counterterms. If we consider
not just isolated integrals, but the procedure of regularization and
subtraction of counterterms as a whole, the limit-interchange can generate
results that differ at most by finite local terms, which is precisely a
scheme change. Once physical normalization conditions are imposed, all
physical quantities coincide.

Moreover, two cut-offs can be identified only up to an arbitrary constant.
For example, we have 
\begin{equation}
\frac{1}{\varepsilon _{2}}=\ln \Lambda _{L}+c,  \label{c}
\end{equation}
and the constant $c$ has no universal meaning. We can even choose different
constants $c$ for each high-energy divergence. Indeed, changing $c$ to $%
c+\delta c$ amounts to shift the pole subtraction from $1/\varepsilon _{2}$
to $1/\varepsilon _{2}-\delta c$ in the high-energy theory. Details about
cut-off identifications are given in Appendix B.

Summarizing, an equivalent regularization of (\ref{lelow}) can be obtained
from LVQED$_{\varepsilon }$, where however the limit $\Lambda
_{L}\rightarrow \infty $ is taken before the limit $\varepsilon
_{2}\rightarrow 0$. When $\Lambda _{L}$ goes to infinity (\ref{le}) just
collapses to (\ref{lelow}). Since $\varepsilon _{2}$ is still non-vanishing,
we just obtain lvQED$_{\varepsilon}$, namely a dimensional regularization of
(\ref{lelow}), where only space is continued to complex dimensions.

Now, one-loop logarithmic divergences are scheme independent, so they can be
calculated removing the two cut-offs in either order. On the other hand,
power-like divergences do depend on the scheme. Since we regard LVQED as a
fundamental theory, not just a partial regularization of (\ref{lelow}), the
powers of $\Lambda _{L}$ must be studied taking $\varepsilon _{2}\rightarrow
0$ first.

It turns out that the power-like divergences in $\Lambda _{L}$ are
multiplied by arbitrary incalculable constants, inherited by the scheme
arbitrariness of the high-energy theory. Thus, they are devoid of any
physical meaning. Ultimately, we discover that it is completely safe to
study the low-energy theory sending $\Lambda _{L}$ to infinity at $%
\varepsilon _{2}\neq 0$.

In the rest of this section we perform a detailed analysis and prove these
statements. A one-loop correlation function is the sum of contributions of
the form $I_{r}/\Lambda _{L}^{r}$, where $r$ is a non-negative integer and 
\begin{equation}
I_{r}=\int \frac{\mathrm{d}\hat{p}\ \mathrm{d}^{3-\varepsilon _{2}}\bar{p}}{%
(2\pi )^{4}}\frac{N_{s}(\hat{p},\bar{p},\hat{k},\bar{k})}{%
\prod_{i=1}^{n}\left[ (\hat{p}-\hat{k}_{i})^{2}+a_{i}(\bar{p}-\bar{k}%
_{i})^{2}+m_{i}^{2}+(\bar{p}-\bar{k}_{i})^{2}\Delta _{i}((\bar{p}-\bar{k}%
_{i})^{2}/\Lambda _{L}^{2})\right] },  \label{gei}
\end{equation}
where $\Delta _{i}(x)$ are some polynomials such that $\Delta _{i}(0)=0$ and
the $k_{i}$'s denote linear combinations of the external momenta $k$. The
numerator $N_{s}$ is a certain monomial of degree $s$ in momenta. Below we
prove that the integral $I_{r}$ is equivalent to 
\begin{equation}
I_{r<}^{\prime }=\int_{|\bar{p}|\leqslant \Lambda _{L}}\frac{\mathrm{d}\hat{p%
}\ \mathrm{d}^{3}\bar{p}}{(2\pi )^{4}}\frac{N_{s}(\hat{p},\bar{p},\hat{k},%
\bar{k})}{\prod_{i=1}^{n}\left[ (\hat{p}-\hat{k}_{i})^{2}+a_{i}(\bar{p}-\bar{%
k}_{i})^{2}+m_{i}^{2}\right] },  \label{geip}
\end{equation}
up to a scheme change, namely up to local counterterms that are at most
power-like divergent. Thus, $I_{r}/\Lambda _{L}^{r}$ is also equivalent to $%
I_{r<}^{\prime }/\Lambda _{L}^{r}$ up to a scheme change. Now, since $%
I_{r<}^{\prime }$ is a one-loop integral, its divergences can only be powers
or logarithms (but not powers times logarithms). By the locality of
counterterms, $I_{r<}^{\prime }$ has the form 
\[
I_{r<}^{\prime }=P(\Lambda _{L},m,k)+P^{\prime }(m,k)\ln \Lambda _{L}+%
\mathrm{finite}+\mathcal{O}(1/\Lambda _{L}), 
\]
where $P$ and $P^{\prime }$ are polynomials. Thus, whenever $r>0$, the
contribution of $I_{r<}^{\prime }/\Lambda _{L}^{r}$ (and $I_{r}/\Lambda
_{L}^{r}$) is just a scheme change. Only the contributions with $r=0$
determine the physical quantities. However, the integrals $I_{0<}^{\prime }$
are precisely those of the low-energy theory regulated with the cut-off $%
\Lambda _{L}$. This proves that the low-energy limit of LVQED can be
studied, up to a scheme change, regulating (\ref{lelow}) with a cut-off $%
\Lambda _{L}$ on the space momenta.

In particular, the scheme-independent contributions to the low-energy
renormalization of LVQED are encoded in $I_{0<}^{\prime }$. Instead, the
scheme-dependent quantities have to be studied directly on LVQED.

The next goal is to prove the equivalence of (\ref{gei}) and (\ref{geip}) up
to a scheme change. As a byproduct, it emerges that the low-energy
power-like divergences are multiplied by arbitrary constants. Before
treating the general case, we illustrate a simple example.

\paragraph{Illustrative example}

Consider the tadpole integral 
\[
I=\int \frac{\mathrm{d}\hat{p}\ \mathrm{d}^{3-\varepsilon _{2}}\bar{p}}{%
(2\pi )^{4}}\frac{1}{D(\hat{p},\bar{p},m)+\bar{p}^{2}\Delta (\bar{p}%
^{2}/\Lambda _{L}^{2})}, 
\]
where 
\[
D(\hat{p},\bar{p},m)=\hat{p}^{2}+a_{2}\bar{p}^{2}+m^{2},\qquad \Delta
(x)=a_{0}x^{2}+a_{1}x, 
\]
and $a_{0},a_{2}>0$. At $\Lambda _{L}$ finite, this integral is
logarithmically divergent. When $\Lambda _{L}\rightarrow \infty $ , it
becomes quadratically divergent.

It is convenient to split the $\bar{p}$-domain of integration in two
regions: the sphere $|\bar{p}|\leqslant \Lambda _{L}$ and the crown $|\bar{p}%
|\geqslant \Lambda _{L}$. Rescaling $\hat{p},\bar{p}$ to $\Lambda _{L}\hat{p}%
,\Lambda _{L}\bar{p}$ we get 
\[
I=I_{<}+I_{>},\qquad I_{\lessgtr }=\Lambda _{L}^{2-\varepsilon _{2}}\int_{|%
\bar{p}|\lessgtr 1}\frac{\mathrm{d}\hat{p}\ \mathrm{d}^{3-\varepsilon _{2}}%
\bar{p}}{(2\pi )^{4}}\frac{1}{D(\hat{p},\bar{p},m/\Lambda _{L})+\bar{p}%
^{2}\Delta (\bar{p}^{2})}. 
\]

We want to show that $I$ is equivalent to 
\[
I_{<}^{\prime }=\int_{|\bar{p}|\leqslant \Lambda _{L}}\frac{\mathrm{d}\hat{p}%
\ \mathrm{d}^{3}\bar{p}}{(2\pi )^{4}}\frac{1}{D(\hat{p},\bar{p},m)}=\Lambda
_{L}^{2}\int_{|\bar{p}|\leqslant 1}\frac{\mathrm{d}\hat{p}\ \mathrm{d}^{3}%
\bar{p}}{(2\pi )^{4}}\frac{1}{D(\hat{p},\bar{p},m/\Lambda _{L})}, 
\]
up to a scheme change.

Consider first $I_{>}$. The integrand can be expanded in powers of $m$
(there are no infrared problems, since $\bar{p}$ cannot approach zero). We
can write 
\[
I_{>}=\sum_{k=0}^{\infty }(-1)^{k}\Lambda _{L}^{2-\varepsilon
_{2}-2k}m^{2k}I_{k},\qquad I_{k}=\int_{|\bar{p}|>1}\frac{\mathrm{d}\hat{p}\ 
\mathrm{d}^{3-\varepsilon _{2}}\bar{p}}{(2\pi )^{4}}\frac{1}{\mathcal{D}%
^{k+1}(\hat{p},\bar{p},0)}. 
\]
where 
\[
\mathcal{D}(\hat{p},\bar{p},0)\equiv D(\hat{p},\bar{p},0)+\bar{p}^{2}\Delta (%
\bar{p}^{2}). 
\]
When $\varepsilon _{2}\rightarrow 0$ only $I_{0}$ diverges. Let us write 
\[
I_{0}=\frac{A_{0}}{\varepsilon _{2}}+B_{0}+\mathcal{O}(\varepsilon
_{2}),\qquad I_{k}=B_{k}+\mathcal{O}(\varepsilon _{2})\quad \mathrm{{\ for}\,%
}k\mathrm{>0,} 
\]
where $A_{i},B_{i}$ are constants. We have, for $\varepsilon _{2}\sim 0$, 
\[
I_{>}=\Lambda _{L}^{2}\left[ A_{0}\left( \frac{1}{\varepsilon _{2}}-\ln
\Lambda _{L}\right) +B_{0}\right] -B_{1}m^{2}+\mathcal{O}(\varepsilon
_{2},m^{2}/\Lambda _{L}^{2}). 
\]
To translate this expression into more familiar terms, just recall that if
we had regulated the high-energy theory with a cut-off $\Lambda $ instead of
using the dimensional regularization, the coefficient of $A_{0}$ between the
square brackets would be ln$(\Lambda /\Lambda _{L})$.

Taking $\Lambda _{L}\rightarrow \infty $ after $\varepsilon _{2}\rightarrow
0 $ we thus find, using (\ref{c}), 
\[
I_{>}\rightarrow \Lambda _{L}^{2}(cA_{0}+B_{0})-B_{1}m^{2}. 
\]
We see that the contribution of the crown does not contain logarithmic
divergences and it is polynomial in the mass. Moreover, the coefficients of
the power-like divergences remain undetermined.

Now, let us study $I_{<}$. Here we can immediately take the limit $%
\varepsilon _{2}\rightarrow 0$, since the integral is UV convergent. Define $%
X$ so that 
\[
I_{<}=I_{<}^{\prime }+\Lambda _{L}^{2}J+m^{2}X, 
\]
where 
\begin{equation}
J=-\int_{|\bar{p}|\leqslant 1}\frac{\mathrm{d}\hat{p}\ \mathrm{d}^{3}\bar{p}%
}{(2\pi )^{4}}\frac{\bar{p}^{2}\Delta (\bar{p}^{2})}{D(\hat{p},\bar{p},0)%
\mathcal{D}(\hat{p},\bar{p},0)}<\infty .  \label{j}
\end{equation}
It is easy to see that $X$ is regular in the limit $\Lambda _{L}\rightarrow
\infty $. Its limit $\bar{X}$ reads 
\[
\bar{X}=\int_{|\bar{p}|\leqslant 1}\frac{\mathrm{d}\hat{p}\ \mathrm{d}^{3}%
\bar{p}}{(2\pi )^{4}}\frac{\bar{p}^{2}\Delta (\bar{p}^{2})\left( D(\hat{p},%
\bar{p},0)+\mathcal{D}(\hat{p},\bar{p},0)\right) }{D^{2}(\hat{p},\bar{p},0)%
\mathcal{D}^{2}(\hat{p},\bar{p},0)}<\infty . 
\]
Here and in (\ref{j}) it is crucial to check the absence of infrared
divergences at $p\sim 0$.

Calculating $I_{<}^{\prime }$ and collecting our results, we get 
\begin{equation}
I=\Lambda _{L}^{2}\left( \frac{1}{8\pi ^{2}a_{2}^{1/2}}+cA_{0}+B_{0}+J%
\right) -m^{2}\left( \frac{\ln (4a_{2}\Lambda _{L}^{2}/m^{2})-1}{16\pi
^{2}a_{2}^{3/2}}+B_{1}-\bar{X}\right) +\mathcal{O}(m^{2}/\Lambda _{L}^{2}).
\label{quadra}
\end{equation}
Thus, the scheme-independent divergences are contained in $I_{<}^{\prime }$.
The quadratic divergences remain arbitrary, due to the constant $c$
inherited from the high-energy theory.

Observe that another argument to justify the identification (\ref{c}) is
that $I$ cannot have divergences of the form $\Lambda _{L}^{2}/\varepsilon
_{2}$ or $\Lambda _{L}^{2}\ln \Lambda _{L}$, because they can arise only at
higher loops.

\paragraph{General case}

Now we give the general argument for the equivalence of (\ref{gei}) and (\ref
{geip}) up to a scheme change. The degree of divergence $\omega $ of $%
I_{r<}^{\prime }$ is $s+4-2n$. If $\omega <0$ the limits $\varepsilon
_{2}\rightarrow 0$ and $\Lambda _{L}\rightarrow \infty $ can be taken
directly on the integrand of $I_{r}$ and the result is equal to the limit $%
\Lambda _{L}\rightarrow \infty $ of $I_{r<}^{\prime }$, which is finite.
Thus, we can assume $\omega \geqslant 0$.

Again, split the $\bar{p}$-domain of integration in two regions: the sphere $%
|\bar{p}|\leqslant \Lambda _{L}$ and the crown $|\bar{p}|\geqslant \Lambda
_{L}$, and call $I_{r>}$ and $I_{r<}$ the two contributions to $I_{r}$.
Rescaling $\hat{p},\bar{p}$ to $\Lambda _{L}\hat{p},\Lambda _{L}\bar{p}$, we
get 
\begin{equation}
I_{r>}=\Lambda _{L}^{\omega -\varepsilon _{2}}\int_{|\bar{p}|\geqslant 1}%
\frac{\mathrm{d}\hat{p}\ \mathrm{d}^{3-\varepsilon _{2}}\bar{p}}{(2\pi )^{4}}%
\frac{N_{s}(\hat{p},\bar{p},\hat{k}/\Lambda _{L},\bar{k}/\Lambda _{L})}{%
\prod_{i=1}^{n}\mathcal{D}_{i}(\hat{p}-\hat{k}_{i}/\Lambda _{L},\bar{p}-\bar{%
k}_{i}/\Lambda _{L},m_{i}/\Lambda _{L})},  \label{irmag}
\end{equation}
where 
\[
\mathcal{D}_{i}(\hat{p},\bar{p},m_{i})=\hat{p}^{2}+a_{i}\bar{p}%
^{2}+m_{i}^{2}+\bar{p}^{2}\Delta _{i}(\bar{p}^{2}). 
\]
Now, expand the expression (\ref{irmag}) in powers of $k/\Lambda _{L}$ and $%
m/\Lambda _{L}$, which is allowed because the integral has an IR cut-off.
After a finite number of terms we get contributions that are finite for $%
\varepsilon _{2}\rightarrow 0$ and disappear when later $\Lambda
_{L}\rightarrow \infty $. Thus the result of these limits on $I_{r>}$ is a
polynomial in $k$ and $m$. The coefficients are powers $\Lambda _{L}^{i}$,
possibly multiplied by simple poles $1/\varepsilon _{2}$. Since 
\[
\frac{\Lambda _{L}^{i-\varepsilon _{2}}}{\varepsilon _{2}}=\Lambda
_{L}^{i}\left( \frac{1}{\varepsilon _{2}}-\ln \Lambda _{L}+\mathcal{O}%
(\varepsilon _{2})\right) \rightarrow \Lambda _{L}^{i}\left( c_{i}+\mathcal{O%
}(\varepsilon _{2})\right) , 
\]
we see that all power-like divergences are multiplied by (different)
arbitrary constants $c_{i}$ and no $\ln \Lambda _{L}$ can appear.

Next, consider $I_{r<}-I_{r<}^{\prime }$. We can set $\varepsilon _{2}=0$,
since there are no ultraviolet divergences here. To keep the notation
simple, let us collect both $k$'s and $m$'s in the same symbol $K$ and leave
index contractions implicit. Define $K^{\omega }X$ as the difference between 
$I_{r<}-I_{r<}^{\prime }$ and its expansion in $k/\Lambda _{L}$ and $%
m/\Lambda _{L}$ up to the order $\omega -1$. We have 
\[
I_{r<}=I_{r<}^{\prime }+\sum_{i=0}^{\omega -1}\Lambda _{L}^{\omega
-i}K^{i}J_{i}+K^{\omega }X. 
\]
Now, by construction all $J_{i}$'s are integrals of functions depending only
on $\hat{p}$ and $\bar{p}$ and no other dimensionful quantities\footnote{%
Here we are talking about the dimensions before the rescaling $\hat{p},\bar{p%
}$ $\rightarrow$ $\Lambda _{L}\hat{p},\Lambda _{L}\bar{p}$.}. Such integrals
have a UV cut-off ($|\bar{p}|\leqslant 1$). Moreover, power counting shows
that they are also IR convergent, because they have dimensions $\omega -i$.
Next, we need to check that the $\Lambda _{L}\rightarrow \infty $ (or $%
K\rightarrow 0$) limit $\bar{X}$ of $X$ is well defined. Again, there are no
UV problems, but we must check IR convergence. Although $X$ has dimension
zero, we must recall that it is originated expanding the difference $%
I_{r<}-I_{r<}^{\prime }$, whose integrand is proportional to a polynomial $%
\Delta (x)=\mathcal{O}(x)$. The factor $\Delta $ enhances the naive IR power
counting by two units, just enough to make $\bar{X}$ well defined.

This concludes the proof.

\section{Low-energy counterterms}

\setcounter{equation}{0}

In this section we compute the renormalization of lvQED. Using the results
of the previous section, we know that we do not need to pay attention to
power-like divergences, so we just focus on the logarithmic ones. The
contributing diagrams are (a), (b), (c), (e), (f), (h) and (i), plus the
same as (h) and (i) but with $\hat{A}$-external legs. The key-integrals are
collected in appendix A. We find 
\begin{eqnarray}
\frac{\Delta _{1}\mathcal{L}_{\mathrm{E}\text{\textrm{-}}\mathrm{low}}}{\ln
(\Lambda _{L}/\mu )} &=&-\frac{e^{2}}{6\pi ^{2}|b_{1}|}\left( \frac{1}{2}F_{%
\hat{\mu}\bar{\nu}}F_{\hat{\mu}\bar{\nu}}+\frac{b_{1}^{2}}{4}F_{\bar{\mu}%
\bar{\nu}}F_{\bar{\mu}\bar{\nu}}\right) -\frac{e^{2}(\tau _{2}+3b_{1}^{2})}{%
4\pi ^{2}|b_{1}|\sqrt{\tau _{2}}(|b_{1}|+\sqrt{\tau _{2}})}m\bar{\psi}\psi 
\\
&&+\frac{e^{2}(\tau _{2}-3b_{1}^{2})}{4\pi ^{2}\sqrt{\tau _{2}}(|b_{1}|+%
\sqrt{\tau _{2}})^{2}}\bar{\psi}\hat{D}\!\!\!\!\slash \psi -\frac{e^{2}(\tau
_{2}+b_{1}^{2})(|b_{1}|+2\sqrt{\tau _{2}})}{12\pi ^{2}|b_{1}|\sqrt{\tau _{2}}%
(|b_{1}|+\sqrt{\tau _{2}})^{2}}b_{1}\bar{\psi}\bar{D}\!\!\!\!\slash \psi .
\end{eqnarray}
Thus, 
\begin{eqnarray*}
\beta _{e} &=&e\gamma _{A}=\frac{e^{3}}{12\pi ^{2}|b_{1}|},\qquad \beta
_{\tau _{2}}=\frac{e^{2}(\tau _{2}-b_{1}^{2})}{6\pi ^{2}|b_{1}|},\qquad
\beta _{b_{1}}=-\frac{e^{2}b_{1}\left( 2|b_{1}|(\tau _{2}-2b_{1}^{2})+\sqrt{%
\tau _{2}}(\tau _{2}+b_{1}^{2})\right) }{6\pi ^{2}|b_{1}|\sqrt{\tau _{2}}%
(|b_{1}|+\sqrt{\tau _{2}})^{2}}, \\
\gamma _{\psi } &=&-\frac{e^{2}(\tau _{2}-3b_{1}^{2})}{8\pi ^{2}\sqrt{\tau
_{2}}(|b_{1}|+\sqrt{\tau _{2}})^{2}},\qquad \beta _{m}=-m\frac{e^{2}(2|b_{1}|%
\sqrt{\tau _{2}}+\tau _{2}+3b_{1}^{2})}{4\pi ^{2}|b_{1}|(|b_{1}|+\sqrt{\tau
_{2}})^{2}}.
\end{eqnarray*}

Around the Lorentz invariant surface our results agree with those found by
Kostelecky, Lane and Pickering \cite{pick}, once restricted to the CPT-, P-
and rotation invariant case. See also the more recent paper \cite{colla}.
Another check of our results is that setting 
\begin{equation}
\tau _{2}=b_{1}^{2},  \label{conda}
\end{equation}
we recover QED. Indeed, when (\ref{conda}) holds, then both $\tau _{2}$ and $%
b_{1}$ can be set to 1 rescaling the space coordinates (as well as the
fields and $e$). Then $\beta _{\tau _{2}}$ and $\beta _{b_{1}}$ vanish,
while $\beta _{e}$, $\gamma _{\psi }$ and $\beta _{m}$ take their known
values.

\section{The new setting of the hierarchy problem}

\setcounter{equation}{0}

In the previous section we have seen that at low energies the power-like
divergences in $\Lambda _{L}$ are multiplied by arbitrary constants, the
arbitrariness being inherited by the divergences of the high-energy theory.
Those arguments are very general, in particular they also apply to the
Lorentz violating Standard Models of \cite{lvsm,noh}. These facts force us
to reconsider the hierarchy problem. For definiteness, we treat the Higgs
mass.

In general, when new physics beyond the Standard Model is assumed, it is
assumed to be described by a finite theory, that contains a physical energy
scale $\Lambda $ and gives the Standard Model when $\Lambda $ is sent to
infinity. Then, at energies much smaller than $\Lambda $ the Higgs mass is
corrected by physical quadratic divergences, and their removal poses a
fine-tuning problem. On the other hand, if the Standard Model were exact at
arbitrarily high-energies, the quadratic divergences of the Higgs mass would
have no physical meaning (among the other things, they would be
scheme-dependent) and could be removed with a mathematical operation devoid
of physical significance.

Our extensions of the Standard Model model do assume new physics beyond the
Standard Model, but not described by a finite theory, rather a
super-renormalizable one. Our results show that the coefficient of the
quadratic divergences is still scheme-dependent and devoid of physical
meaning. In this section we explain that, because of this, no fine-tuning
problem arises. We stress that our statement does not contraddict the common
lore about the hierarchy problem, because our models do not obey the
finiteness assumption.

The general form of the (one-loop) mass renormalization can be read for
example from (\ref{quadra}). We have 
\begin{equation}
m_{\Lambda }^{2}=m^{2}+a\Lambda _{L}^{2}\ln \frac{\Lambda ^{2}}{\Lambda
_{L}^{2}}+bm^{2}\ln \frac{\Lambda _{L}^{2}}{m^{2}}+c\Lambda _{L}^{2}+dm^{2}.
\label{erg}
\end{equation}
Here $m_{\Lambda }$ denotes the bare mass, $m$ is the low-energy mass, $%
\Lambda $ is the ultraviolet cut-off (we have replaced $1/\varepsilon _{2}$
with $\ln \Lambda +$constant), while $a$, $b$ and $d$ are calculable
coefficients, depending on the parameters of the theory. In\ LVQED the
formula of the electron-mass renormalization has a form analogous to (\ref
{erg}), but the squares $m_{\Lambda }^{2}$, $m^{2}$, $\Lambda ^{2}$ and $%
\Lambda _{L}^{2}$ are replaced by $m_{\Lambda }$, $m$, $\Lambda $ and $%
\Lambda _{L}$, respectively, and the coefficient $a$ can be read from (\ref
{formu}).

If $\Lambda $ were the physical scale introduced by a finite ultraviolet
completion of the theory, $c$ would also be calculable. Then we would have a
fine-tuning problem: roughly, $m^{2}$ is small and $a\Lambda _{L}^{2}\ln
(\Lambda ^{2}/\Lambda _{L}^{2})$ is large, so $m_{\Lambda }^{2}$ is also
large and 
\[
m^{2}=\text{small}=\text{large}-\text{large.} 
\]
On the other hand, if our models are regarded as fundamental models of the
Universe (when gravity is switched off), namely if we assume that no more
fundamental models exist beyond them, then $\Lambda $ is an unphysical
cut-off, which means that it must be sent to infinity, and $c$ remains
scheme-dependent, therefore arbitrary. Then, both $a\Lambda _{L}^{2}\ln
(\Lambda ^{2}/\Lambda _{L}^{2})$ and $m_{\Lambda }^{2}$ are infinite, so 
\[
m^{2}=\infty -\infty . 
\]
This cancellation between infinities is just the usual job of
renormalization. There is no fine-tuning problem, because $m^{2}$ cannot be
said to be small or large with respect to infinity.

We can make this even clearer eliminating the cut-off $\Lambda $. Formula (%
\ref{erg}) incorporates also the (one-loop) running from energies $\Lambda $
to energies $\Lambda _{L}$. In other words, if we substitute $\Lambda $ with 
$\Lambda _{L}$ formula (\ref{erg}) gives an expression for the Higgs mass $%
m_{L}$ at the scale of Lorentz violation. We find 
\[
m_{L}^{2}=m^{2}+bm^{2}\ln \frac{\Lambda _{L}^{2}}{m^{2}}+c\Lambda
_{L}^{2}+dm^{2}. 
\]
We see that the quadratic divergence $\sim \Lambda _{L}^{2}$ is still
multiplied by the meaningless arbitrary constant $c$, which cannot be
eliminated. There is no reason why the quantity $c\Lambda _{L}^{2}$ should
be large, even if $\Lambda _{L}^{2}$ is large. Actually, we can use the
arbitrariness of $c$ to make it disappear, and obtain 
\[
m_{L}^{2}=m^{2}+bm^{2}\ln \frac{\Lambda _{L}^{2}}{m^{2}}+dm^{2}. 
\]
Again, we do not find any fine-tuning problem.

Our argument is very general. It does not depend on the particular
high-energy completion of the theory, as long as it is not finite. Indeed,
if the UV completion is not finite, at some point we do need an unphysical
cut-off $\Lambda $, which brings some arbitrariness into the game and makes
the quadratic divergences unphysical.

In conclusion, the hierarchy problem is a true problem only if the ultimate
theory of the Universe is completely finite. If the ultimate theory of the
Universe is just renormalizable, or even super-renormalizable, for example
one of the models that we propose, then the hierarchy problem disappears.

\section{Conclusions}

\setcounter{equation}{0}

In this paper we have studied the one-loop renormalization of high-energy
Lorentz violating QED, a subsector of the Lorentz violating Extended
Standard Model proposed recently. We have also analyzed the interplay
between high-energy and low-energy renormalizations in detail.

We have shown that the high-energy theory leaves important remnants at
low energies, such as incalculable, arbitrary factors in front of all
power-like divergences. This property holds under the sole assumption that
the fundamental theory beyond the Standard Model, whether it is (\ref{noH})
or not, is not completely finite, but just renormalizable, or even
super-renormalizable. In particular, the arbitrariness inherited by the
high-energy theory allows us to eliminate the quadratically divergent
corrections to the Higgs mass, thereby removing the hierarchy problem.

\vskip 20truept \noindent {\Large \textbf{Acknowledgments}}

\vskip 10truept

One of us (D.A.) is grateful to D. Buttazzo for useful discussions.

\vskip 20truept \noindent {\Large \textbf{Appendix A: Key integrals}}

\vskip 10truept

\renewcommand{\theequation}{A.\arabic{equation}} \setcounter{equation}{0}

For the calculations of the high-energy renormalization we just need the
divergent part of one integral, namely 
\[
\int \frac{\mathrm{d}\hat{p}\ \mathrm{d}^{3-\varepsilon _{2}}\bar{p}}{(2\pi
)^{4}}\frac{\hat{p}^{q}(\bar{p}^{2})^{r}(\bar{p}\cdot \bar{k})^{s}}{\left( 
\hat{p}^{2}+b_{0}^{2}\frac{(\bar{p}^{2})^{3}}{\Lambda _{L}^{4}}+\delta
^{2}\right) ^{k_{1}}\left( \hat{p}^{2}+\tau _{0}\frac{(\bar{p}^{2})^{3}}{%
\Lambda _{L}^{4}}+\delta ^{2}\right) ^{k_{2}}},
\]
for $2+q+(s+2r)/3=2(k_{1}+k_{2})$. Using Feynman parameters we can
immediately integrate over $\hat{p}$. This isolates the pole of the $\bar{p}$%
-integral, therefore the divergent part. The remaining integral over the
Feynman parameter gives a hypergeometric function. The final result is 
\begin{eqnarray*}
&&\frac{\tau _{0}^{(1+q)/2-k_{1}-k_{2}}}{\Lambda _{L}^{2+2q-4k_{1}-4k_{2}}}%
\frac{(\bar{k}^{2})^{s/2}}{\varepsilon _{2}}\frac{\left( 1+(-1)^{s}\right)
\left( 1+(-1)^{q}\right) \Gamma \left( \frac{q+1}{2}\right) \Gamma \left(
k_{1}+k_{2}-\frac{q+1}{2}\right) }{2(s+1)(2\pi )^{3}\Gamma (k_{1}+k_{2})}%
\times  \\
&&\qquad \qquad \qquad \times \ _{2}F_{1}\left( k_{1},k_{1}+k_{2}-\frac{q+1}{%
2},k_{1}+k_{2},1-\frac{b_{0}^{2}}{\tau _{0}}\right) .
\end{eqnarray*}

For the calculations of the low-energy renormalization we need the
logarithmic divergences of two integrals, namely 
\begin{equation}
\int_{|\bar{p}|\leqslant \Lambda _{L}}\frac{\mathrm{d}\hat{p}\ \mathrm{d}^{3}%
\bar{p}}{(2\pi )^{4}}\frac{(\hat{p}^{2},\bar{p}^{2})}{\left( \hat{p}%
^{2}+b_{1}^{2}\bar{p}^{2}+m^{2}\right) ^{2}\left( \hat{p}^{2}+\tau _{2}\bar{p%
}^{2}\right) }\sim \frac{\ln (\Lambda _{L}/m)}{8\pi ^{2}|b_{1}|(|b_{1}|+%
\sqrt{\tau _{2}})^{2}}\left( 1,\frac{2|b_{1}|+\sqrt{\tau _{2}}}{b_{1}^{2}%
\sqrt{\tau _{2}}}\right) .  \label{biss}
\end{equation}
As usual, the one-loop calculation is done expanding in external momenta.
This gives a sum of contributions involving the integrals (\ref{biss}), plus
more standard integrals and integrals that do not have logarithmic
divergences.

\vskip 20truept \noindent {\Large \textbf{Appendix B: Identification of
cut-offs}}

\vskip 10truept

\renewcommand{\theequation}{B.\arabic{equation}} \setcounter{equation}{0}

Formula (\ref{c}) can be proved comparing two different regularizations of
the same integral. The first technique is a dimensional regularization where
only the space dimension is continued to complex values. The second
technique is a higher-derivative regularization where only higher-space
derivatives are used. We get 
\begin{eqnarray*}
\int \frac{\mathrm{d}\hat{p}\ \mathrm{d}^{3-\varepsilon _{2}}\bar{p}}{(2\pi
)^{4}}\frac{1}{(\hat{p}^{2}+\bar{p}^{2}+m^{2})^{2}} &=&\frac{2}{(4\pi
)^{2}\varepsilon _{2}}+\mathrm{constant,} \\
\int \frac{\mathrm{d}\hat{p}\ \mathrm{d}^{3}\bar{p}}{(2\pi )^{4}}\frac{1}{%
\left( \hat{p}^{2}+\bar{p}^{2}+m^{2}+\frac{(\bar{p}^{2}+m^{2})^{2}}{\Lambda
_{L}^{2}}\right) ^{2}} &=&\frac{\ln (\Lambda _{L}/m)}{8\pi ^{2}}+\mathrm{%
constant,}
\end{eqnarray*}
whence (\ref{c}) follows. Similarly, if we use a cut-off on the $\bar{p}$%
-integral instead of higher-space derivatives, we get 
\[
\int_{|\bar{p}|\leqslant \Lambda _{L}}\frac{\mathrm{d}\hat{p}\ \mathrm{d}^{3}%
\bar{p}}{(2\pi )^{4}}\frac{1}{\left( \hat{p}^{2}+\bar{p}^{2}+m^{2}\right)
^{2}}=\frac{\ln (\Lambda _{L}/m)}{8\pi ^{2}}+\mathrm{constant.}
\]

\end{document}